\documentclass[12pt]{article}
\usepackage{amsfonts,amsmath,graphicx,setspace}
\usepackage{color,hyperref,verbatim,natbib,rotating}

\definecolor{Red}{rgb}{0.5,0,0}
\definecolor{Blue}{rgb}{0,0,0.5}
\hypersetup{%
    colorlinks = {true},
    linktocpage = {true},
    plainpages = {false},
    linkcolor = {Blue},
    citecolor = {Blue},
    urlcolor = {Red},
    pdfstartview = {XYZ null null 1.25},
    pdfpagemode = {UseOutlines},
    pdfview = {XYZ null null null}
}
\newcommand{\email}[1]{\href{mailto:#1}{\normalfont\texttt{#1}}}
\newcommand{\bfalpha}{\mbox{{\boldmath $\alpha$}}}
\newcommand{\bfbeta}{\mbox{{\boldmath $\beta$}}}
\newcommand{\bfpsi}{\mbox{{\boldmath $\psi$}}}
\newcommand{\bfgamma}{\mbox{{\boldmath $\gamma$}}}
\newcommand{\bftheta}{\mbox{{\boldmath $\theta$}}}
\newcommand{\bflambda}{\mbox{{\boldmath $\lambda$}}}
\newcommand{\bfdelta}{\mbox{{\boldmath $\delta$}}}

\newcommand{\bw}{\mbox{{\boldmath $w$}}}

\newcommand{\by}{\mbox{{\boldmath $y$}}}
\newcommand{\bx}{\mbox{{\boldmath $x$}}}

\newcommand{\bz}{\mbox{{\boldmath $z$}}}

\newcommand{\bY}{\mbox{{\boldmath $Y$}}}
\newcommand{\bT}{\mbox{{\boldmath $T$}}}
\newcommand{\bb}{\mbox{{\boldmath $b$}}}

\newcommand{\dd}{\mbox{d}}
\newcommand{\bD}{\mbox{{\boldmath $D$}}}
\newcommand{\bK}{\mbox{{\boldmath $K$}}}

\def\argmin{\mathop{\rm arg\,min}}%

\newcommand{\EPCE}{E$\widehat{\mbox{PC}}$E}
\newcommand\independent{\protect\mathpalette{\protect\independenT}{\perp}}
\def\independenT#1#2{\mathrel{\rlap{$#1#2$}\mkern2mu{#1#2}}}

\bibpunct{(}{)}{;}{a}{}{,}

\begin{document}

{\vspace*{1cm}}

\begin{center}
\Large \bf Optimizing Dynamic Predictions from Joint Models using Super Learning
\end{center}

\vspace{0.5cm}

\begin{center}
{\large Dimitris Rizopoulos$^{1,2,*}$ and Jeremy M.G. Taylor$^{3}$}\footnote{$^*$Correspondence at: Department of Biostatistics, Erasmus University Medical Center, PO Box 2040, 3000 CA Rotterdam, the Netherlands. E-mail address: \email{d.rizopoulos@erasmusmc.nl}.}\\
$^{1}$Department of Biostatistics, Erasmus University Medical Center, the Netherlands\\
$^{2}$Department of Epidemiology, Erasmus University Medical Center, the Netherlands\\
$^{3}$Department of Biostatistics, University of Michigan, Ann Arbor, USA\\
\end{center}

\begin{spacing}{1}
\noindent {\bf Abstract}\\
Joint models for longitudinal and time-to-event data are often employed to calculate dynamic individualized predictions used in numerous applications of precision medicine. Two components of joint models that influence the accuracy of these predictions are the shape of the longitudinal trajectories and the functional form linking the longitudinal outcome history to the hazard of the event. Finding a single well-specified model that produces accurate predictions for all subjects and follow-up times can be challenging, especially when considering multiple longitudinal outcomes. In this work, we use the concept of super learning and avoid selecting a single model. In particular, we specify a weighted combination of the dynamic predictions calculated from a library of joint models with different specifications. The weights are selected to optimize a predictive accuracy metric using V-fold cross-validation. We use as predictive accuracy measures the expected quadratic prediction error and the expected predictive cross-entropy. In a simulation study, we found that the super learning approach produces results very similar to the Oracle model, which was the model with the best performance in the test datasets. All proposed methodology is implemented in the freely available \textsf{R} package \textbf{JMbayes2}.
\end{spacing}

\vspace{0.8cm}


\section{Introduction} \label{Sec:Introduction}
Joint models for longitudinal and time-to-event data have been established as a versatile tool for calculating dynamic predictions for longitudinal and survival outcomes \citep{taylor.et.al:05, rizopoulos:11, taylor.et.al:13}. The advantageous feature of these predictions is that they are updated over time as extra information becomes available. As a result, they have found numerous applications in precision medicine, including cancer and cardiovascular diseases.

The motivation for our research comes from prostate cancer patients who, after diagnosis, underwent surgical removal of the prostate gland (radical prostatectomy). The treating physicians closely monitor the prostate-specific antigen (PSA) levels of these patients to determine the risk of recurrence and metastasis. Increasing PSA values suggest the cancer may be regrowing, although it is generally not yet detectable on imaging. After the initial surgery, PSA levels drop to near zero; however, PSA may rise again for some patients, leading the treating physicians to recommend salvage therapy to reduce their risk of metastasis. After salvage therapy, PSA levels nearly always drop, sometimes substantially, but typically rise again if metastasis is going to occur.

Optimizing the accuracy of dynamic predictions from joint models is a difficult task. In particular, previous research has shown that two aspects are important \citep{ferrer.et.al:19}. First, the time trend specification in the mixed-effects models, and second, the functional forms that specify how the longitudinal histories are linked to the hazard of the event. Previous applications of joint models have considered a single model for obtaining dynamic predictions. However, due to the aforementioned complexities, finding a well-specified model can be challenging, especially when multiple longitudinal outcomes are considered. Moreover, due to the dynamic nature of these predictions, different models may provide different levels of predictive accuracy at different follow-up times.

Standard approaches for model selection within the Bayesian framework include information criteria and (pseudo-) Bayes factors. However, these methods provide an overall assessment of a model's fit. In contrast, we are particularly interested in the model that best predicts future events given that subject $i$ was event-free up to time point $t$, i.e., the conditional cumulative risk probabilities. Moreover, in practice, it will seldom be the case that a single model provides the most accurate predictions for all subjects and time points. In this work, we consider multiple joint models instead and combine the dynamic predictions from these models to optimize the predictive accuracy. Under the Bayesian paradigm, an established method to combine quantities from different statistical models is Bayesian model averaging (BMA) \citep{hoeting.et.al:99}. \cite{rizopoulos.et.al:14} have previously used BMA to combine dynamic predictions from a library of joint models. Even though BMA enjoys several optimality properties, it also has some disadvantages. First, it requires calculating the marginal likelihood under each posited model, which is a computationally demanding task. Second, the BMA weights are not explicitly designed to optimize the accuracy of predictions, and it is unclear if they appropriately account for over-fitting. Finally, the weights are defined using the posterior distribution of each model given the data. As a result, even though the competing models may not differ much in their posterior densities (considering the likelihood's magnitude), the differences are large enough that often, one model dominates the weights over the others.

We will use the concept of super learning (SL) to overcome these issues \citep{breiman:96, vanderlaan.et.al:06, naimi.balzer:18, phillips.et.al:23}. SL is an ensemble method that allows researchers to combine several different prediction algorithms into one where the candidate algorithms can be quite different. It uses $V$-fold cross-validation to build the optimally weighted combination of predictions from a library of candidate algorithms. Optimality is defined by a user-specified objective function, such as minimizing the mean squared error or maximizing the area under the receiver operating characteristic curve. The theoretical properties of the super learning procedure have been established by \cite{vanderlaan.et.al:07}. Here we will consider a library of joint models with different specifications and present appropriate objective functions for optimizing the accuracy of dynamic predictions. In particular, we focus on different formulations of the time effect for the longitudinal outcome and different functional forms to link this outcome with the event process. We measure the dynamic predictions' accuracy using the expected Brier score and the expected predictive cross-entropy and show how these are formulated under the SL framework.

The rest of the paper is organized as follows: Section~\ref{Sec:JMs} presents the joint models' framework. Section~\ref{Sec:SuperLearn} introduces the Super Learning technique in the context of joint models and dynamic predictions and presents the two accuracy measures for combining the predictions from different models. Section~\ref{Sec:UMPD_Analysis} employs SL in the context of the University of Michigan Prostatectomy Data, and Section~\ref{Sec:Simulation} shows the results of a simulation study that evaluates the performance of SL. Finally, Section~\ref{Sec:Discussion} closes the paper with a discussion.


\section{Joint Modeling Framework} \label{Sec:JMs}
\subsection{Joint Models} \label{Sec:JMs_def}
In this section, we present a general definition of the joint modeling framework for longitudinal and time-to-event data \citep{rizopoulos:12}. Let $\mathcal D_n = \{T_i, \delta_i, \by_i; i = 1, \ldots, n\}$ denote a sample from the target population, where $T_i^*$ denotes the true event time for the $i$-th subject, $C_i$ the censoring time, $T_i = \min(T_i^*, C_i)$ the corresponding observed event time, and $\delta_i = \mathbb{I}(T_i^* \leq C_i)$ the event indicator, with $\mathbb{I}(\cdot)$ being the indicator function that takes the value 1 when $T_i^* \leq C_i$, and 0 otherwise. In addition, we let $\by_i$ denote the $n_i \times 1$ longitudinal response vector for the $i$-th subject, with element $y_{il}$ denoting the value of the longitudinal outcome taken at time point $t_{il}$, $l = 1, \ldots, n_i$.

To accommodate different types of longitudinal responses in a unified framework, we postulate that the response vector $\by_i$ conditional on the vector of unobserved random effects $\bb_i$ has a distribution $\mathcal F_\psi$ parameterized by the vector $\bfpsi$. This more general formulation allows for distributions not covered by the exponential family. The mean of the distribution of the longitudinal outcome conditional on the random effects has the form
\[
g \bigl [ E \{ y_i(t) \mid \bb_i \} \bigr ] = \eta_i(t) =
\bx_i^\top(t) \bfbeta + \bz_i^\top(t) \bb_i,
\]
where $g(\cdot)$ denotes a known one-to-one monotonic link function, and $y_i(t)$ denotes the value of the longitudinal outcome for the $i$-th subject at time point $t$, $\bx_i(t)$ and $\bz_i(t)$ denote the time-dependent design vectors for the fixed-effects $\bfbeta$ and for the random effects $\bb_i$, respectively. We let $\phi$ denote the scale parameter of $\mathcal F_\psi$, i.e., $\bfpsi^\top = (\bfbeta^\top, \phi)$. The random effects are assumed to follow a multivariate normal distribution with mean zero and variance-covariance matrix $\bD$. For the survival process, we assume that the risk of an event depends on a function of the subject-specific linear predictor $\eta_i(t)$ and the random effects. More specifically, we have
\begin{eqnarray*}
h_i \{t \mid \mathcal H_i(t), \bw_i\} & = &
\lim_{s \rightarrow 0} \Pr \{ t \leq T_i^* < t + s \mid T_i^* \geq t, \mathcal H_i(t),
\bw_i \} \big / s\\
& = & h_0(t) \exp \bigl  [\bfgamma^\top
\bw_i + f \{\eta_i(t), \bb_i, \bfalpha \} \bigr] , \quad t > 0,
\end{eqnarray*}
where $\mathcal H_i(t) = \{ \eta_i(s), 0 \leq s < t \}$ denotes the history of the underlying longitudinal process up to $t$, $h_0(\cdot)$ denotes the baseline hazard function, $\bw_i$ is a vector of baseline covariates with corresponding regression coefficients $\bfgamma$. Finally,  the baseline hazard function $h_0(\cdot)$ is modeled flexibly using a B-splines approach, i.e.,
\[
\log h_0(t) = \sum \limits_{p = 1}^P \gamma_{h_0,p} B_p(t, \bflambda),
\]
where $B_p(t, \bflambda)$ denotes the $p$-th basis function of a B-spline with knots $\lambda_1, \ldots, \lambda_P$ and $\bfgamma_{h_0}$ the vector of spline coefficients.

The function $f(\cdot)$, parameterized by vector $\bfalpha$, specifies which components/features of the longitudinal outcome process are included in the linear predictor of the relative risk model. Some examples previously considered in the literature are \citep{rizopoulos:12, rizopoulos.et.al:14, Papageorgiou.et.al:19}:
\begin{eqnarray*}
f \{\mathcal H_i(t), \bb_i, \bfalpha \} = \left \{
\begin{array}{l}
\alpha \eta_i(t),\\
\alpha \eta_i'(t),
\mbox{ with } \eta_i'(t) = \frac{\dd \eta_i(t)}{\dd t},\\
\alpha \eta_i''(t),
\mbox{ with } \eta_i''(t) = \frac{\dd^2\eta_i(t)}{\dd t^2},\\
\alpha \displaystyle \frac{1}{v} \displaystyle \int_{t - v}^t \eta_i(s) \, \dd s, \;\; 0 < v \leq t,\\
\bfalpha^\top \bb_i.
\end{array}
\right.
\end{eqnarray*}
These formulations of $f(\cdot)$ postulate that the hazard of an event at time $t$ is associated with the underlying level of the biomarker at the same time point, the slope/velocity of the biomarker at $t$, the acceleration of the biomarker at $t$, the average biomarker level in the period $(t - v, t)$, or the random effects alone. Combinations of these functional forms and their interactions with baseline covariates are also often considered.

An advantage of the joint modeling framework is that it provides valid inferences for the joint distribution $\{T_i^*, \by_i\}$ even if the censoring process depends on the observed longitudinal history and the covariates included in the model, without requiring to additionally model this process \citep{tsiatis.davidian:04}. The same property also holds for the visiting process, i.e., the process determining the time points at which the longitudinal measurements are collected.


\subsection{Bayesian Estimation} \label{Sec:JMs_estimation}
We let $\bftheta^\top = (\bfbeta^\top, \phi, \bfgamma_{h_0}^\top, \bfgamma^\top, \bfalpha^\top, \mbox{vech}(\bD)^\top, \tau)$ to denote the vector of all model parameters, where $\mbox{vech}(\bD)$ denotes the unique elements of the variance-covariance matrix $\bD$. We proceed under the Bayesian paradigm and draw inferences using the joint posterior distribution $\{\bftheta, \bb \mid \bY, \bT, \bfdelta\}$, where $\bY$, $\bT$, and $\bfdelta$ denote outcome vectors for all $n$ subjects. We use standard priors for $\bftheta$, i.e., normal priors for all regression coefficients $(\bfbeta, \bfgamma, \bfgamma_{h_0}, \bfalpha)$, inverse-Gamma priors for $\phi$ and the diagonal elements of $\bD$, and the LKJ prior for the correlation matrix of the random effects \citep{lewandowski.et.al:09}. To ensure smoothness of the baseline hazard function $h_0(t)$, we postulate a `penalized' prior distribution for the regression coefficients $\bfgamma_{h_0}$:
\[
p(\bfgamma_{h_0} \mid \tau) \propto \tau^{\rho(K)/2}\exp \Bigl (-\frac{\tau}{2}
\bfgamma_{h_0}^\top \bK \bfgamma_{h_0} \Bigr ),
\]
where $\tau$ is the smoothing parameter that takes a $\mbox{Gamma}(5, 0.05)$ hyper-prior in order to ensure a proper posterior for $\bfgamma_{h_0}$, $\bK = \Delta_r^\top \Delta_r$, where $\Delta_r$ denotes $r$-th difference penalty matrix, and $\rho(\bK)$ denotes the rank of $\bK$. We should note here that $\tau$ is also included in $\bftheta$.

We use a Markov chain Monte Carlo (MCMC) approach to obtain samples from the posterior distribution for all model parameters and the random effects. This algorithm is implemented in the freely available \textsf{R} package \textbf{JMbayes2} \citep{JMbayes2} that we used to fit the model.


\section{Optimizing Predictions via Super Learning} \label{Sec:SuperLearn}
\subsection{Model Weights} \label{Sec:SuperLearn_def}
The basic idea behind super learning is to derive model weights that optimize the cross-validated predictions. More specifically, we let $\mathcal{L} = \{M_1, \ldots, M_L\}$ denote a library with $L$ models. There are no restrictions to the models included in this library, and actually, it is recommended to consider a wide range of possible models. Among others, these joint models differ in the specification of the time trend in the longitudinal submodels (e.g., linear or nonlinear trajectories), the functional form for the longitudinal outcome in the event submodel, and the functional form of the other covariates (e.g., interactions and nonlinear terms) in both submodels.

We split the original dataset $\mathcal{D}_n$ in $V$ folds. The choice of $V$ will depend on the size and number of events in $\mathcal{D}_n$ \citep{phillips.et.al:23}. In particular, for each fold, we need to have a sufficient number of events to robustly quantify the predictive performance. Using the cross-validation method, we fit the $L$ models in the combined $v-1$ folds, and we will calculate predictions for the $v$-th fold we left outside. Due to the dynamic nature of the predictions, we want to derive optimal weights at different follow-up times. More specifically, we consider the sequence of time points $t_1, \ldots, t_Q$. The number and placing of these time points should again consider the available event information in $\mathcal{D}_n$. For example, we should have at least 10-15 events per time interval $(t_{q-1}, t_q)$, with $q = 1, \ldots, Q$, and $t_0 = 0$.

For any $t_q \in \{t_1, \ldots, t_Q\}$, we define $\mathcal{R}(t_q, v)$ to denote the subjects at risk at time $t_q$ that belong to the $v$-th fold. For all subjects in $\mathcal{R}(t_q, v)$, we calculate the cross-validated predictions (conditioning on the covariates $\bw_i$, $\bx_i$ and $\bz_i$ is assumed in the following expressions but is dropped to simplify notation),
\[
\hat{\pi}_i^{(v)}(t_q + \Delta t \mid t_q, M_l) = \Pr \{T_i^* < t_q + \Delta t \mid T_i^* > t_q, \mathcal H_i(t), M_l, \mathcal{D}_n^{(-v)}\}.
\]
These predictions are calculated based on model $M_l$ in library $\mathcal{L}$ that was fitted in the dataset $\mathcal{D}_n^{(-v)}$ that excludes the subjects in the $v$-th fold. The calculation is based on a Monte Carlo approach \citep{rizopoulos:11}. We define $\hat{\tilde{\pi}}_i^{v}(t_q + \Delta t \mid t_q)$ to denote the convex combination of the $L$ predictions, i.e.,
\[
\hat{\tilde{\pi}}_i^{v}(t_q + \Delta t \mid t_q) = \sum\limits_{l = 1}^L \varpi_l(t_q) \hat{\pi}_i^{(v)}(t_q + \Delta t \mid t_q, M_l), \quad \mbox{for all } v \in {1, \ldots, V},
\]
with $\varpi_l(t_q) > 0$, for $l = 1, \ldots, L$, and $\sum_l \varpi_l(t_q) = 1$. Note that the weights $\varpi_l(\cdot)$ are time-varying, i.e., at different follow-up times, different combinations of the $L$ models may yield more accurate predictions. The weighted combination of the predictions from the $L$ models is typically called the ensemble super learner (eSL). The model with the best cross-validated prediction metric is called the discrete super learner (dSL). Most often, but not always, this is the model with the largest weight $\varpi_l(\cdot)$.


\subsection{Measuring Predictive Performance} \label{Sec:SuperLearn_Brier}
For the ensemble super learner and for any time $t$, we will select the weights $\{\varpi_l(t); l = 1, \ldots, L\}$ that optimize the predictive performance of the combined cross-validated predictions. We will employ the framework of proper scoring rules to achieve that. A scoring rule $\mathcal{S}\{\pi_i(u \mid t), \mathbb{I}(t < T_i^* < u)\}$ is called proper if the true distribution achieves the optimal expected score, i.e., in our case if
\[
E \Bigl [\mathcal{S}\{\pi_i^{true}(u \mid t), \; \mathbb{I}(t < T_i^* < u) \} \Bigr] \leq E \Bigl [\mathcal{S}\{\hat{\pi}_i(u \mid t), \; \mathbb{I}(t < T_i^* < u) \} \Bigr], \quad u > t,
\]
where $\pi_i^{true}(u \mid t)$ denotes the conditional risk probabilities under the true model, and $\hat{\pi}_i(u \mid t)$ is an estimate of $\pi_i^{true}(u \mid t)$. The expectation is taken with respect to the conditional density of the survival outcome under the true model $\{T_i^* \mid T_i^* > t, \mathcal Y_i(t)\}$, with $\mathcal Y_i(t) = \{ y_i(t_{il}); 0 \leq t_{il} \leq t, l = 1, \ldots, n_i\}$, and the scoring rule $\mathcal S(\cdot, \cdot)$ is defined such that a lower score indicates better accuracy.

The Brier score is a proper scoring rule that combines discrimination and calibration to measure overall predictive performance. In particular, at follow-up time $t$ and for a medically-relevant time window $\Delta t$, we define the Brier score as
\[
\mbox{BS}(t + \Delta t, t) = E \biggl [\Bigl \{ \mathbb{I}(T_i^* \leq t + \Delta t) - \tilde{\pi}_i^{v}(t + \Delta t \mid t) \Bigr\}^2 \; \Big | \; T_i^* > t \biggr].
\]
The Brier score evaluates the predictive accuracy at time $t + \Delta t$. To summarize the predictive accuracy in the interval $(t, t + \Delta t]$, we can use the integrated Brier score,
\[
\mbox{IBS}(t + \Delta t, t) = \frac{1}{\Delta t} \int_{t}^{t + \Delta t} \mbox{BS}(s, t) \; \dd s.
\]
In practice, this integral does not have a closed-form solution and needs to be approximated numerically. In our examples, we use Simpson's rule to approximate it, i.e.,
\[
\mbox{IBS}(t + \Delta t, t) \approx \frac{2}{3} \mbox{BS}(t + 0.5\Delta t, t) + \frac{1}{6} \mbox{BS}(t + \Delta t, t),
\]
where we have dropped the first term of the approximation because $\mbox{BS}(t, t) = 0$.

To estimate $\mbox{BS}(t + \Delta t, t)$, we need to appropriately account for patients who were censored in the interval $(t, t + \Delta t]$. Two approaches to achieve this is using inverse probability of censoring weights (IPCW), and model-based weights. In the former approach, we only use the subjects who had an event before $t + \Delta t$, and the ones for whom we know that they survived up to $t + \Delta t$; patients censored in $(t, t + \Delta t]$ do not contribute \citep{blanche.et.al:15}, that is,
\[
\widehat{\mbox{BS}}_{IPCW}(t + \Delta t, t) = \frac{1}{n_t} \sum \limits_{i: T_i > t} \widehat{W}_i(t + \Delta t, t) \Bigl \{ \mathbb{I}(T_i \leq t + \Delta t) - \hat{\tilde{\pi}}_i^{v}(t + \Delta t \mid t) \Bigr\}^2,
\]
where $n_t$ denotes the number of subjects at risk at time $t$, and
\[
\widehat{W}_i(t + \Delta t, t) = \frac{\mathbb{I}(t < T_i \leq t + \Delta t) \delta_i}{\hat{G}(T_i \mid t)} + \frac{\mathbb{I}(T_i > t + \Delta t)}{\hat{G}(t + \Delta t \mid t)} + 0 \cdot \mathbb{I}(t < T_i \leq t + \Delta t) (1 - \delta_i),
\]
with $\hat{G}(\cdot)$ denoting Kaplan-Meier estimate of the censoring distribution $\Pr(C_i > t)$.

The model-based approach uses all subjects at risk at time $t$ \citep{henderson.et.al:02}, i.e.,
\begin{eqnarray*}
\lefteqn{\widehat{\mbox{BS}}_{model}(t + \Delta t, t) = }\\
&& \frac{1}{n_t} \sum \limits_{i: T_i > t} \delta_i  \mathbb{I}(T_i \leq t + \Delta t) \Bigl \{1 - \hat{\tilde{\pi}}_i^{v}(t + \Delta t \mid t) \Bigr\}^2 + \mathbb{I}(T_i > t + \Delta t) \Bigl \{\hat{\tilde{\pi}}_i^{v}(t + \Delta t \mid t) \Bigr\}^2\\
&& \; + \; (1 - \delta_i) \mathbb{I}(T_i \leq t + \Delta t) \biggl [ \hat{\tilde{\pi}}_i^{v}(t + \Delta t \mid T_i) \Bigl \{1 - \hat{\tilde{\pi}}_i^{v}(t + \Delta t \mid t) \Bigr\}^2\\
&& \quad\quad + \; \Bigl \{1 - \hat{\tilde{\pi}}_i^{v}(t + \Delta t \mid T_i) \Bigr \}\Bigl \{\hat{\tilde{\pi}}_i^{v}(t + \Delta t \mid t) \Bigr\}^2\biggr].
\end{eqnarray*}

Both approaches have their merits and flaws. The IPCW approach is model-free in the sense that there is no assumption about the correctness of the specification of the joint model used for computing $\hat{\pi}_i^{(v)}(t_q + \Delta t \mid t_q, M_l)$. On the other hand, it makes the strong assumption that censoring does not depend on covariates and, more importantly, on the history of the longitudinal biomarker $\mathcal Y_i(t)$. This assumption is crucial because the Brier score is a proper scoring rule only when it holds \citep{rindt.et.al:22}. When an inaccurate assumption is made for the censoring weights, we may achieve an incorrect Brier score that is lower in value (i.e., better) than the Brier score under the true weights \citep{kvamme.borgab:23}. In the clinical and epidemiological contexts where joint models are applied, physicians may decide to remove patients from a study based on their previously observed longitudinal measurements. Moreover, it may also be the case that different physicians use in another way the longitudinal history to exclude patients, e.g., one physician may decide to use just the last observed value of the biomarker. In contrast, another may remove patients who showed a sudden increase/decrease in the biomarker in the last two measurements. In situations like this, specifying an appropriate model for the censoring process is challenging, and thus, the appealing features of the IPCW approach are lost.

The model-based weights approach does not require modeling the censoring process. In particular, it will provide valid estimates of the Brier score even when censoring depends in a complex manner on the longitudinal history. However, it requires that the joint model is well-specified, which, as argued in Section~\ref{Sec:Introduction}, is also a strong assumption. And when we compare two or more models, it is paradoxical that all of them will be correctly specified, leading to unfair comparisons between them. However, in the context we consider here, the disadvantage of the model-based approach is alleviated. This is because we are not interested in comparing between models but rather combining the predictions from these models to improve accuracy. That is, the weights for the censored subjects are not calculated from a single model but as a convex combination of the weights from the $L$ models. In addition, as we will see later, the weights $\{\varpi_l(t); l = 1, \ldots, L\}$ are derived from the cross-validated predictions that ensure that predictions from over-fitted models will be down-weighted.

As an alternative proper scoring rule in the interval $(t, t + \Delta t]$ we consider an adaptation of the expected predictive cross-entropy proposed by \cite{commenges.et.al:12}:
\[
\mbox{EPCE}(t + \Delta t, t) = E \biggl \{-\log \Bigl [ p \bigl \{ T_i^* \mid t < T_i^* \leq t + \Delta t, \mathcal Y_i(t), \mathcal D_{n} \bigr \} \Bigr ] \biggr \},
\]
where the expectation is taken with respect to $\{T_i^* \mid T_i^* > t, \mathcal Y_i(t)\}$ under the true model. An estimate of $\mbox{EPCE}(t + \Delta t, t)$ that accounts for censoring can be obtained using the sample at hand,
\[
\mbox{\EPCE}(t + \Delta t, t) = \frac{1}{n_t} \sum \limits_{i: T_i > t} - \log \Bigl [ p \bigr \{\tilde T_i, \tilde \delta_i \mid T_i > t, \mathcal Y_i(t), \mathcal D_n \bigl \} \Bigr ], \]
where $\tilde T_i = \min(T_i, t + \Delta t)$ and $\tilde \delta_i = \delta_i \mathbb{I}(t < T_i \leq t + \Delta t)$. The $\mbox{EPCE}(t + \Delta t, t)$ makes the same assumptions regarding censoring as the model-based weights for $\mbox{BS}(t + \Delta t, t)$; namely that $T_i^* \independent C_i \mid \mathcal Y_i(t), \bw_i$. However, because no model weights are required to account for censoring, it does not share the disadvantage of $\mbox{BS}(t + \Delta t, t)$ mentioned above. The disadvantage of the $\mbox{EPCE}(t + \Delta t, t)$ over the Brier score is interpretability. Namely, the Brier score estimates a well-defined parameter in the population, the mean squared distance between the observed and expected outcomes. The square root of the Brier score is thus the expected distance between the observed and predicted value on the probability scale. On the contrary, the value of the expected predictive cross-entropy does not have a simple interpretation.

To use the EPCE for obtaining the super-learning model-specific weights, we need to formulate it as a function of the dynamic predictions from a joint model. It is convenient to redefine $\pi_i(u \mid t, M_l)$ as the dynamic subject-specific survival probabilities, i.e.,
\[
\pi_i(u \mid t, M_l) = \Pr \{T_i^* > u \mid T_i^* > t, \mathcal Y_i(t), \mathcal D_n, M_l\}, \quad u > t.
\]
Then we write the conditional predictive log-likelihood as (conditioning on $M_l$ is assumed but is omitted from the following expressions for exposition):
\[
\log \Bigl [ p \bigr \{\tilde T_i, \tilde \delta_i \mid T_i > t, \mathcal Y_i(t), \mathcal D_n \bigl \} \Bigr] = \tilde \delta_i \log [ h_i \{\tilde T_i \mid \mathcal Y_i(t), \mathcal D_n\} ] + \log \frac{\Pr \{T_i^* > \tilde T_i \mid \mathcal Y_i(t), \mathcal D_n\}}{\Pr \{T_i^* > t \mid \mathcal Y_i(t), \mathcal D_n\}}.
\]
The second term is $\log \{\pi_i(\tilde T_i \mid t)\}$. For the first term, we write the hazard function as
\[
h_i \{\tilde T_i \mid \mathcal Y_i(t), \mathcal D_n\} = - \frac{\frac{\dd}{\dd \displaystyle t} \Pr \{T_i^* > t \mid \mathcal Y_i(t), \mathcal D_n\} \Big |_{t = \tilde T_i}}{\Pr \{T_i^* > \tilde T_i \mid \mathcal Y_i(t), \mathcal D_n\}} ,
\]
and we approximate the derivative with a forward difference, i.e.,
\begin{eqnarray*}
h_i \{\tilde T_i \mid \mathcal Y_i(t), \mathcal D_n\} & \approx & - \frac{\Pr \{T_i^* > \tilde T_i + \epsilon \mid \mathcal Y_i(t), \mathcal D_n\} - \Pr \{T_i^* > \tilde T_i \mid \mathcal Y_i(t), \mathcal D_n\} }{\epsilon \Pr \{T_i^* > \tilde T_i \mid \mathcal Y_i(t), \mathcal D_n\}}\\
& = & \frac{1 - \pi_i(\tilde T_i + \epsilon \mid \tilde T_i)}{\epsilon}, \quad \epsilon \rightarrow 0.
\end{eqnarray*}
Combining these two terms, we get the final expression:
\begin{eqnarray*}
\mbox{\EPCE}(t + \Delta t, t) = -\frac{1}{n_t} \sum \limits_{i: T_i > t} \tilde \delta_i \bigl [\log \{1 - \pi_i(\tilde T_i + \epsilon \mid \tilde T_i)\} - \log(\epsilon) \bigr ] + \log \{\pi_i(\tilde T_i \mid t)\}.
\end{eqnarray*}
In practice, we can compute $\mbox{\EPCE}(t + \Delta t, t)$ using a small value for $\epsilon$, e.g., $\epsilon = 0.001$. Numerical experiments we performed showed that the EPCE values are minimally affected by the value of $\epsilon$. In settings where a better approximation is required, the central difference approximation can be used, i.e.,
\begin{eqnarray*}
h_i \{\tilde T_i \mid \mathcal Y_i(t), \mathcal D_n\} & \approx & - \frac{\Pr \{T_i^* > \tilde T_i + \epsilon \mid \mathcal Y_i(t), \mathcal D_n\} - \Pr \{T_i^* > \tilde T_i - \epsilon \mid \mathcal Y_i(t), \mathcal D_n\} }{2\epsilon \Pr \{T_i^* > \tilde T_i \mid \mathcal Y_i(t), \mathcal D_n\}},
\end{eqnarray*}
which has a truncation error on the order of $\epsilon^2$ but requires an extra function evaluation.\citep{press.et.al:07}

In our context, both $\mbox{BS}(t + \Delta t, t)$ and $\mbox{EPCE}(t + \Delta t, t)$ are calculated using the convex combination of the cross-validated predictions $\hat{\tilde{\pi}}_i^{v}(t + \Delta t \mid t)$. In particular, using the super-learning procedure, we obtain the weights $\widehat{\varpi}_l(t)$ that minimize a proper scoring rule (in our case, either the $\mbox{BS}(t + \Delta t, t)$ or $\mbox{EPCE}(t + \Delta t, t)$) of the cross-validated predictions,
\begin{eqnarray*}
\widehat{\varpi}_l(t) & = & \argmin_{\varpi} \biggl [ \sum \limits_{v = 1}^V \mathcal{S} \Bigl \{ \sum_{l = 1}^L \varpi_l \hat{\pi}_i^{(v)}(t + \Delta t \mid t, M_l), T_i, \delta_i \Bigr \} \biggr],
\end{eqnarray*}
under the constraints $\varpi_l(t) > 0$, for $l = 1, \ldots, L$, and $\sum_l \varpi_l(t) = 1$. We can transform to an unconstrained optimization problem using the logistic transformation and use a general-purpose minimization algorithm (e.g., using functions \texttt{optim()} or \texttt{nlminb()} in \textsf{R}). The vignette \textit{`Combined Dynamic Predictions via Super Learning'} (available at \url{https://drizopoulos.github.io/JMbayes2/articles/Super_Learning.html}) describes how the super learning procedure is implemented in package \textbf{JMbayes2}.


\subsection{Specification of the Models Library} \label{Sec:SuperLearn_Library}
A consideration to improve the performance of the super learning procedure is to make an informed selection of the models to include in library $\mathcal L$. Any such model pre-selection should be done carefully to ensure an objective assessment of the procedure's predictive performance. In particular, a data-driven selection of models to include in $\mathcal L$ would need to be validated to avoid over-optimistic accuracy measures. Because the cross-validation procedure in the super-learning algorithm is used to obtain the cross-validated predictions from all members of $\mathcal L$, the validation of a data-driven selection of the library would need to add another layer of re-sampling. For example, a Bootstrap procedure to populate $\mathcal L$ with a nested cross-validation procedure to combine the predictions for the models selected in the specific Bootstrap sample. Even though such an approach is technically possible, two challenges for its implementation are that the data-driven models' selection needs to be automatic (e.g., not using figures to assess proportional hazards but a test), and the computational burden would increase drastically.


\section{University of Michigan Prostatectomy Data Analysis} \label{Sec:UMPD_Analysis}
We return to our motivating University of Michigan Prostatectomy Data. This database includes 3634 PCa patients who underwent a radical prostatectomy during the period 1996--2013. Of those patients, 271 (15.6\%) received salvage therapy, 102 (2.8\%) developed metastasis, and 209 (5.8\%) died without metastases. Of these 209 patients, 190 died before salvage therapy, and 19 died after. Since death due to prostate cancer is extremely rare without prior metastases, these deaths are considered due to other causes. Here, we focus on metastasis. PSA levels have been routinely collected for these patients to monitor progression. There are a total of 24673 PSA measurements, with a median of six measurements per patient (IQR: six measurements). We aim to utilize the longitudinal PSA measurements and baseline information (Gleason score, T-stage of the tumors, age, race, comorbidities) to predict metastasis and aid the decision-making process of urologists. Our analysis assumes that the decision to initiate salvage therapy at a time point $t$ may depend on the observed PSA history before $t$ and baseline covariates. In the context of causal inference in the presence of time-varying confounders, this corresponds to the sequential exchangeability assumption. Because joint models provide a complete specification of the joint distribution $\{T_i^*, \by_i\}$, they provide valid metastasis predictions without explicitly modeling the salvage treatment assignment mechanism.

We considered four versions of the linear mixed-effects model for the log(PSA + 1) longitudinal outcome. In the first model, we specified linear subject-specific time trends for log(PSA + 1) that change after the first salvage therapy. The second model considers the same specification as the previous model, but we additionally include the baseline covariates age at surgery, Charlson's index, Gleason score, and baseline PSA. These covariates are allowed to have a different effect after salvage. The third model considers nonlinear subject-specific time trends before salvage and linear subject-specific time trends after salvage. The final fourth model has the same specification for the time trends as the third one, but we again include the same covariates as in the second model. Using each of these linear mixed models, we fitted three joint models with different specifications of the hazard submodel for metastasis. In each hazard model, we include the covariates mentioned in the second and fourth linear mixed models above and a time-varying component. Each hazard submodel has a different specification of the time-varying component. In the first specification, we consider the current value of log(PSA + 1); in the second one, the current value and the velocity of log(PSA + 1); and in the third one, the mean log(PSA + 1) from the start of the follow-up. Each of these specifications has two branches, one before and one after salvage therapy. Hence, in total, we consider twelve joint models. The exact formulation of these models is presented in the supplementary material.

We split the UMP data into five folds and fitted the twelve joint models, holding out a fold each time. We calculated the cross-validated predictions from these models for the fold not used when fitting them. We aim to evaluate the predictive performance of the twelve models in two medically relevant time intervals $(t, t + \Delta t]$, namely, $(4, 7]$ and $(6, 9]$. At follow-up year 4, 2514 patients were still at risk, and 28 patients had metastasis in the interval $(4, 7]$. At follow-up year 6, 1914 patients were still at risk, and 16 patients had metastasis in the interval $(6, 9]$. An issue in assessing the predictive accuracy in these intervals is the potential change in salvage therapy status for some patients in the interval $(t, t + \Delta t]$. More specifically, there are three scenarios for the patients at risk at $t$: (i) a patient received salvage before $t$, (ii) a patient did not receive salvage up to $t + \Delta t$, and (iii) a patient received salvage after $t$ and before $t + \Delta t$. For Scenarios (i) and (ii), the predictions are calculated from the model using the correct salvage therapy status in $(t, t + \Delta t]$. However, for Scenario (iii), we calculate predictions as in Scenario (ii), namely, assuming that the patient did not initiate salvage by $t + \Delta t$. Hence, our assessment of the model's predictive accuracy follows the \emph{intention-to-treat} principle, as if patients were randomized at $t$ to receive salvage therapy or not, and we do not change the group in which a patient was `randomized' even if he changed group after $t$. We use the integrated Brier score and the expected predictive cross-entropy as predictive performance metrics. We calculated these metrics for each model using its cross-validated predictions. We also obtained the super learning weights that combined the dynamic predictions from these models to optimize the two prediction metrics and calculated the two metrics using the weighted predictions. Tables~\ref{Tab:IBS} and \ref{Tab:EPCE} present the results.
\begin{table}[ht]
\centering
\caption{Integrated Brier Score (IBS) for the University of Michigan Prostatectomy Data under the twelve joint models, and their combination using super learning. Results are based on 5-fold cross-validation.} \label{Tab:IBS}
\begin{tabular}{rrrlrr}
  \hline
  & \multicolumn{2}{c}{$(t, t + \Delta t] = (4, 7]$} & & \multicolumn{2}{c}{$(t, t + \Delta t] = (6, 9]$}\\
 & IBS & weights &   & IBS & weights \\
  \hline
SL & 0.00639 &  &  & 0.00530 &  \\
  linear-noCov-value & 0.00627 & 0.08334 &  & 0.00538 & 0.00352 \\
  linear-noCov-slope & 0.00631 & 0.08342 &  & 0.00546 & 0.00136 \\
  linear-noCov-mean & 0.00649 & 0.08329 &  & 0.00530 & 0.31705 \\
  linear-Cov-value & 0.00626 & 0.08344 &  & 0.00538 & 0.01732 \\
  linear-Cov-slope & 0.00630 & 0.08342 &  & 0.00547 & 0.00124 \\
  linear-Cov-mean & 0.00652 & 0.08327 &  & 0.00529 & 0.52883 \\
  nonlinear-noCov-value & 0.00646 & 0.08331 &  & 0.00552 & 0.00009 \\
  nonlinear-noCov-slope & 0.00644 & 0.08332 &  & 0.00558 & 0.00002 \\
  nonlinear-noCov-mean & 0.00653 & 0.08327 &  & 0.00535 & 0.08450 \\
  nonlinear-Cov-value & 0.00643 & 0.08332 &  & 0.00553 & 0.00008 \\
  nonlinear-Cov-slope & 0.00643 & 0.08333 &  & 0.00552 & 0.00008 \\
  nonlinear-Cov-mean & 0.00654 & 0.08326 &  & 0.00537 & 0.04589 \\
   \hline
\end{tabular}
\end{table}
\begin{table}[ht]
\centering
\caption{Expected predictive cross-entropy (EPCE) for the University of Michigan Prostatectomy Data under the twelve joint models, and their combination using super learning. Results are based on 5-fold cross-validation.} \label{Tab:EPCE}
\begin{tabular}{rrrlrr}
  \hline
  & \multicolumn{2}{c}{$(t, t + \Delta t] = (4, 7]$} & & \multicolumn{2}{c}{$(t, t + \Delta t] = (6, 9]$}\\
 & EPCE & weights &   & EPCE & weights \\
  \hline
SL & 0.07208 &  &  & 0.05166 &  \\
  linear-noCov-value & 0.07347 & 0.00026 &  & 0.05543 & 0.03259 \\
  linear-noCov-slope & 0.07299 & 0.00004 &  & 0.05471 & 0.15417 \\
  linear-noCov-mean & 0.07476 & 0.00235 &  & 0.05365 & 0.01329 \\
  linear-Cov-value & 0.07338 & 0.00000 &  & 0.05506 & 0.02167 \\
  linear-Cov-slope & 0.07298 & 0.00006 &  & 0.05455 & 0.09639 \\
  linear-Cov-mean & 0.07484 & 0.00274 &  & 0.05353 & 0.02836 \\
  nonlinear-noCov-value & 0.07324 & 0.00000 &  & 0.05376 & 0.01562 \\
  nonlinear-noCov-slope & 0.07242 & 0.79539 &  & 0.05436 & 0.00317 \\
  nonlinear-noCov-mean & 0.07457 & 0.18346 &  & 0.05303 & 0.02524 \\
  nonlinear-Cov-value & 0.07316 & 0.00000 &  & 0.05337 & 0.10840 \\
  nonlinear-Cov-slope & 0.07265 & 0.00121 &  & 0.05283 & 0.08131 \\
  nonlinear-Cov-mean & 0.07454 & 0.01448 &  & 0.05284 & 0.41979 \\
   \hline
\end{tabular}
\end{table}
We observed that for both time intervals the differences between the integrated Brier score are small. However, in the $(6, 9]$ interval the IBE for the joint model with linear subject-specific time trends and the mean log(PSA + 1) functional form (with and without baseline covariates) is smaller than the other model specifications. This is also reflected in the corresponding model weights. In particular, the model weights for the IBE in the interval $(4, 7]$ are effectively equal for all models, but in the $(6, 9]$ interval, the joint model specification mentioned above dominates the weights distribution. The expected predictive cross-entropy seems more sensitive in quantifying differences in the predictive performance of the different models. This also results in smaller super learning estimates of the EPCE than in each of the twelve models, and for both time intervals. In the $(4, 7]$ interval the nonlinear joint models with no covariates and the slope and mean functional form dominate the weights. In the $(6, 9]$ interval the weights are distributed among almost all models.


\section{Simulation Study} \label{Sec:Simulation}
\subsection{Objectives and Design} \label{Sec:Simulation_Desing}
We have performed a simulation study to evaluate the performance of the super learning algorithm in the joint modeling context. For the longitudinal submodel, we consider two settings for the shape of the subject-specific longitudinal trajectories, namely, linear and nonlinear. For the survival submodel, we consider three settings for the functional forms, namely, the underlying value, the velocity, and the average longitudinal outcome level in the period $(0, t)$. Hence, in total, we consider six data-generating models. We simulated training and testing data under three scenarios; in Scenario I, each simulated training dataset contains 125 subjects from each of the six models ($n = 750$); in Scenario II, each simulated training dataset contains 450 subjects from the nonlinear model using the velocity functional form, and 60 subjects from each of the five other models ($n = 750$). Scenarios III has the same settings as Scenario I, but the censoring mechanism is informative. That is, the probability of withdrawing a subject from the training dataset depends on their longitudinal outcome. The testing datasets under all scenarios have the same specification as the corresponding training datasets. The censoring was for all scenarios, on average, 45\%. The parameter values and the other simulation settings we used can be found in the supplementary material. For each scenario, we simulated 250 datasets.

For each simulated training dataset, we implemented three-fold cross-validation, fitting the six models described above. The cross-validated predictions from each model were used to estimate the model weights to optimize predictive performance via the super learning procedure. The discrete super learner was also compared with the ensemble super learner. As predictive performance metrics, we used the integrated Brier score and the EPCE in two time intervals $(t, t + \Delta t]$, namely, $(16, 18]$ and $(19, 21]$. In Scenarios I and II, we used Kaplan-Meier weights to account for censoring; in Scenario III, we used both Kaplan-Meier and model-based weights.


\subsection{Results} \label{Sec:Simulation_Results}
\begin{figure}
\centering{\includegraphics[width=\textwidth]{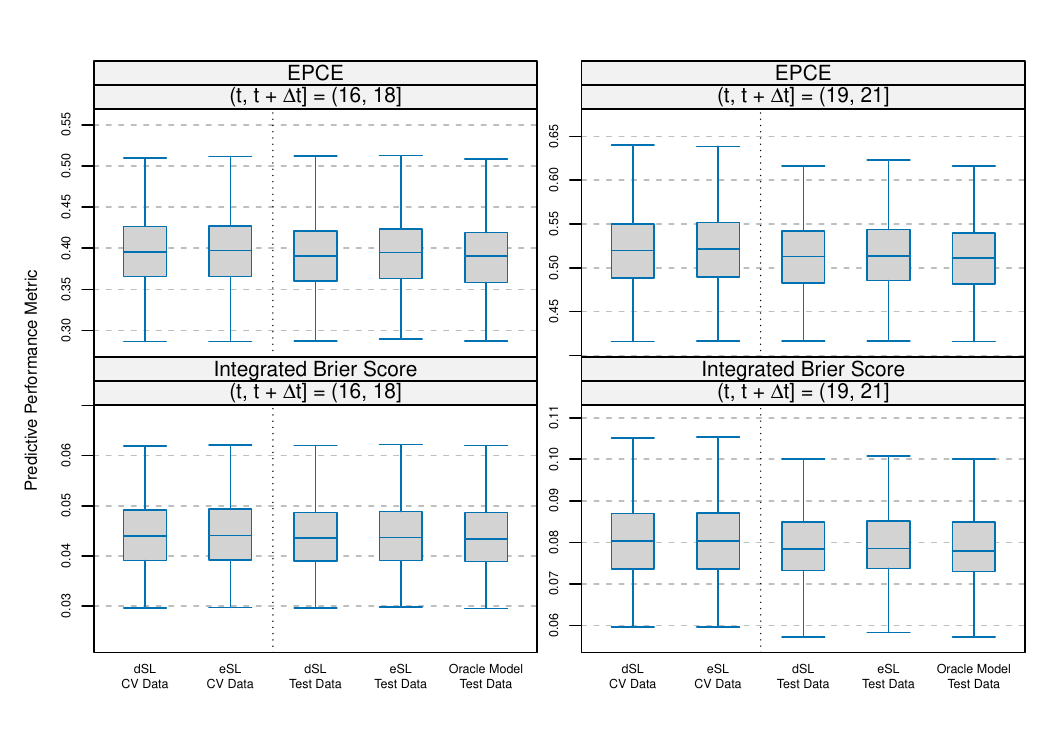}}
\caption{Simulation results for the expected predictive cross-entropy (EPCE) and the integrated Brier score under Scenario I. The left and right panels show the results for the two time intervals $(t, t + \Delta t]$. In each panel, five boxplots are shown summarizing the results over the 250 simulated datasets; the first two show the results of the discrete super learner (dSL) and the ensemble super learner (eSL) from the cross-validation procedure in the training dataset; the last three show the results of dSL, eSL, and the oracle model in the independent testing dataset. Each simulated training and testing dataset contains 750 subjects.} \label{Fig:Sc1}
\end{figure}

\begin{figure}
\centering{\includegraphics[width=\textwidth]{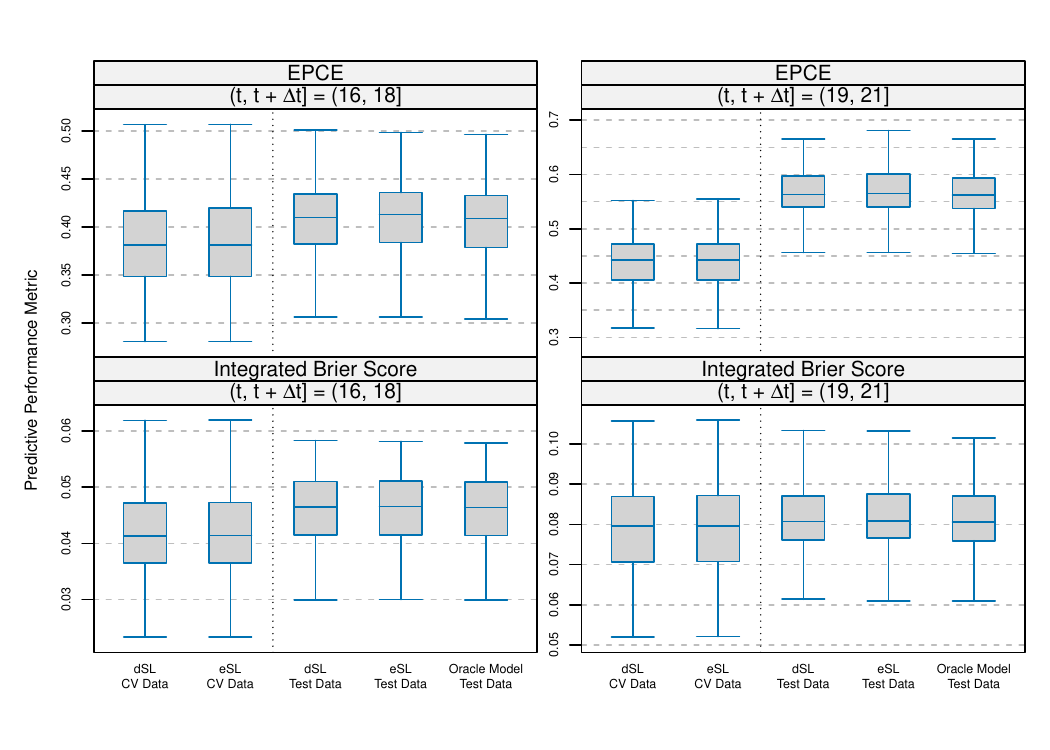}}
\caption{Simulation results for the expected predictive cross-entropy (EPCE) and the integrated Brier score under Scenario II. The left and right panels show the results for the two time intervals $(t, t + \Delta t]$. In each panel, five boxplots are shown summarizing the results over the 250 simulated datasets; the first two show the results of the discrete super learner (dSL) and the ensemble super learner (eSL) from the cross-validation procedure in the training dataset; the last three show the results of dSL, eSL, and the oracle model in the independent testing dataset. Each simulated training and testing dataset contains 750 subjects.} \label{Fig:Sc2}
\end{figure}

\begin{figure}
\centering{\includegraphics[width=\textwidth]{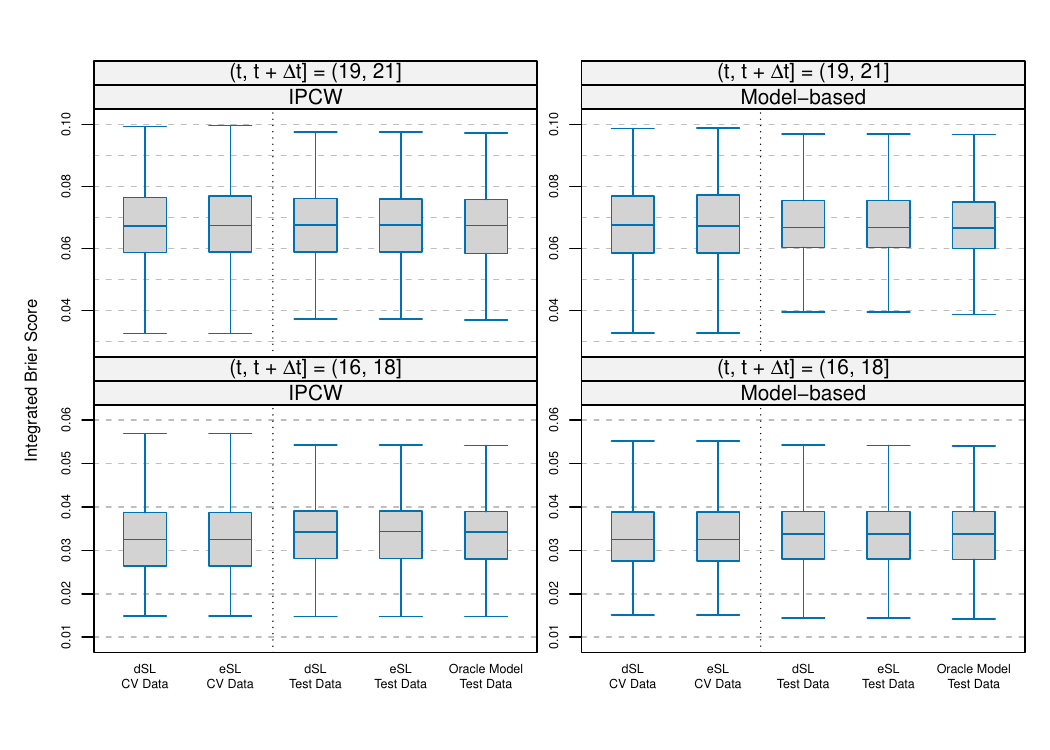}}
\caption{Simulation results for the integrated Brier score under Scenario III. The top and bottom panels show the results for the two time intervals $(t, t + \Delta t]$. The left and right panels show the results using model-based and inverse probability of censoring weights to account for informative censoring. In each panel, five boxplots are shown summarizing the results over the 250 simulated datasets; the first two show the results of the discrete super learner (dSL) and the ensemble super learner (eSL) from the cross-validation procedure in the training dataset; the last three show the results of dSL, eSL, and the oracle model in the independent testing dataset. Each simulated training and testing dataset contains 750 subjects.} \label{Fig:Sc3}
\end{figure}

Figures~\ref{Fig:Sc1} and \ref{Fig:Sc2} show the results for the integrated Brier score and the EPCE under the first two scenarios, respectively.
In each figure, the left and right panels show the results for the two time intervals $(16, 18]$ and $(19, 21]$, respectively. The top and bottom panels show the results for the two accuracy metrics. In each panel, five boxplots are depicted summarizing the results over the 250 simulated datasets. The first boxplot shows the results of the corresponding predictive performance metric for the discrete super learner (dSL), i.e., the model that performed the best in the cross-validation procedure applied in the training dataset. The second boxplot corresponds to the result of the ensemble super learner (eSL) in the training dataset using the cross-validated predictions. The third and fourth boxplots show the results under the dSL and eSL, respectively, but are evaluated in the independent testing dataset. The last boxplot shows the results of the Oracle model, i.e., the model with the lowest value for the predictive performance metric in the testing dataset. Figure~\ref{Fig:Sc3} shows the results for the integrated Brier score under Scenario III.
The left and right panels correspond to the model-based and inverse probability of censoring weights to account for the informative censoring. The top and bottom panels show the results for the two time intervals $(19, 21]$ and $(16, 18]$, respectively. In each panel, the same five boxplots are depicted as in Figures~\ref{Fig:Sc1}--\ref{Fig:Sc2}.

Both metrics indicate that the predictive accuracy in the interval $(19, 21]$ is worse than in $(16, 18]$. This is attributed to the fact that more events are observed in the $(16, 18]$ interval than in the $(19, 21]$ one, leading to a better estimate of the underlying baseline hazard function. In the cross-validated training datasets, eSL performs as well as the dSL under both Scenarios I and II. The performance of the dSL and eSL is the same as that of the oracle model in the independent test datasets. Figures~3 and 4 in the supplementary material zoom in on the performance of the dSL and eSL relative to the oracle model in the test datasets. From these figures, we observe that dSL performs slightly better than the eSL. In Scenario II, all metrics perform worse in the test datasets than in the cross-validated training datasets, indicating a greater degree of overfitting. In Scenario I, both metrics perform slightly better in the test dataset than in the cross-validated training ones. Tables~1 and 2 in the supplementary material show the median estimated weight and interquartile range per model, predictive metric and interval for the first two scenarios, respectively. We see that in Scenario I all models receive the same weight for both metrics and intervals. In Scenario II, the weights for the EPCE metric favor the models with the velocity functional form and the nonlinear time trajectories. The weights for IBS in Scenario II do not favor any specific model. In Scenario III, we observe small differences between the IPCW and the model-based weights to account for the informative censoring. We also observe that the IBS and EPCE select different discrete super learners. Formally, both measures are proper scoring rules, i.e., they are minimized for the predictions derived from the true model that describes the data-generating mechanism. However, this does not mean they should behave the same when comparing different misspecified models. \cite{merkle.steyvers:13} provide a formal comparison between proper scoring rules, which showcases that conclusions vary greatly across different rules so that one's choice of scoring rule should be informed by the forecasting domain.


\section{Discussion} \label{Sec:Discussion}
In this paper, we presented an adaptation of the super learning framework for optimizing dynamic predictions from joint models for longitudinal and time-to-event data. We considered the Brier score and the expected predictive cross-entropy as predictive accuracy metrics and compared two super learning versions. Namely, the discrete super learner selects the model with the best cross-validated prediction metric among the candidate models, and the ensemble super learner specifies an optimal convex combination of the predictions from all candidate models. In the University of Michigan Prostatectomy Data, the ensemble super learner performed better than the discrete one, especially when using the expected predictive cross-entropy as an accuracy metric. In the simulation study both versions performed as well as the oracle model selector; however, the discrete super learned performed slightly better than the ensemble one (Figures~3 and 4 in the supplementary material). Therefore, following the recommendation of \citep{phillips.et.al:23} we also recommend that the eSL is evaluated as another candidate in a dSL, as this allows for the eSL's cross-validated risk to be compared with that of each learner considered in its library. If the eSL performs better than any other candidate, the dSL will select the eSL. Regarding the two predictive accuracy metrics, the performance of the expected predictive cross-entropy was better than the one of the Brier score. The suboptimal performance of the super learner for the Brier score can be explained by the limited ability of this metric to discern between models \citep{wilks:10, benedetti:10}.

Another popular prediction metric is the area under the receiver operating characteristic curve (AUC). The AUC has a more clinically desirable interpretation as the probability to correctly classify patients as `cases' and `controls'. However, we have selected not to include the AUC as a scoring rule to combine predictions from joint models for two reasons. First, the AUC is not a proper scoring rule \citep{rindt.et.al:22}. That is, the AUC is not maximized for the predictions derived from the true model that describes the data-generating mechanism. We find this to be an important property for deriving the weights that optimize the combined predictions from the different models in the library we consider. A related issue with the AUC is that it is scale-invariant and does not measure whether the predicted probabilities are accurate; for example, if we divide the predicted probabilities by a constant, the AUC remains unchanged. Second, the AUC is included in our developments as one of the Brier score components. Namely, the Brier score can be decomposed into two terms, the first measuring the calibration, and the second the discrimination as summarized by the AUC.

A disadvantage of the super learning procedure is that it sacrifices interpretability for predictive performance. In particular, by combining the predictions from the different models, we cannot identify which features of the longitudinal process (described by the different functional forms presented in Section~\ref{Sec:JMs_def}) are most predictive of the risk of an event unless the models we consider differ only in the specification of the functional form and no other aspect. In such a case, the weights we derive to combine the predictions from the various models would tell us which functional forms are most predictive.

The use of $V$-fold cross-validation is key for optimizing the performance of super learners. However, this requirement makes SL a computationally challenging task. We utilized parallel computing to mitigate the computational cost in both the analysis of the University of Michigan Prostatectomy Data and the simulation. Another practical aspect of using $V$-fold cross-validation is the choice of $V$. For binary outcomes, \cite{phillips.et.al:23} suggested using larger values of $V$ for a smaller effective sample size. In our test simulation runs, we found that smaller values of $V$ produce more stable results for smaller effective sample sizes. This is because in the time-to-event data context, we apply cross-validation at different follow-up times, and we require to have a sufficient number of events in each of these intervals. In this setting, selecting a larger $V$ would result in time intervals having very few or no events.

The super learning framework we presented here can be augmented with alternative procedures for dynamic predictions; for instance, the landmark approach \citep{vanhouwelingen.putter:11} or machine learning procedures, such as deep learning \citep{lee.et.al:20}. However, these approaches do not account for informative censoring. To account for this setting, an adaptation would be required by, e.g., specifying a model for the censoring process and incorporating inverse probability of censoring weights. Similarly, these approaches would require further adaptations to work in the University of Michigan Prostatectomy Data and account for the assignment of salvage therapy based on the PSA history. For these reasons, we have not included them in the library of models from which we derive dynamic predictions. In addition, the SL framework could be extended in settings with competing risks and multi-state processes or recurrent events. However, such extensions require appropriate adaptations of the predictive accuracy metrics.

\section*{Funding}
The authors thank the NIH CISNET Prostate Award CA253910 for financial support.

\section*{Supplementary Material}
Supplementary material are available with this paper.


\bibliographystyle{biom}
\bibliography{Functional_Forms.bib}

\end{document}